\documentclass{aa}
\usepackage{graphicx}

\newcommand{\ergscm}{erg\,s$^{-1}$\,cm$^{-2}$}

\begin{document}
   \title{Short-term emission line and continuum variations in Mrk\,110}

   \author{W. Kollatschny
          \inst{1,2}\fnmsep
          \thanks{Based on observations obtained with the Hobby-Eberly
               Telescope, which is a joint project of the University
               of Texas at Austin, the Pennsylvania State
           University, Stanford University, Ludwig-Maximilians-Universit\"at
            M\"unchen, and Georg-August-Universit\"at G\"ottingen.},
          K. Bischoff\inst{1}, E.L. Robinson\inst{2},
          W.F. Welsh\inst{2,3}, and G.J. Hill\inst{2} 
          }

   \offprints{W. Kollatschny}

   \institute{Universit\"{a}ts-Sternwarte G\"{o}ttingen,
              Geismarlandstra{\ss }e 11, D-37083 G\"{o}ttingen, Germany\\
              \email{wkollat@uni-sw.gwdg.de}
         \and
             Department of Astronomy and McDonald Observatory,
             University of Texas at Austin, Austin, TX 78712, USA
         \and
             Department of Astronomy, San Diego State University,
             San Diego, CA 92182, USA}

   \date{Received date, 2001; accepted date, 2001}

   \abstract{We present results of a
    variability campaign of Mrk\,110 performed
    with the 9.2-m Hobby-Eberly Telescope (HET) at McDonald Observatory.
    The high S/N spectra cover most of the optical range. They
    were taken from 1999 November through 
    2000 May. The average interval
    between the observations was 7.3 days and the median interval was only
    3.0 days.
    Mrk\,110 is a narrow-line Seyfert 1 galaxy. 
    During our campaign the continuum flux was in a
    historically low stage.
    Considering the delays of the emission lines with respect to the
    continuum variations we could verify  
    an ionization stratification of the 
    BLR. We derived virial masses of the central     
    black hole from the radial distances of the different emission lines
    and from their widths. 
    The calculated central masses agree within 20\%.
   Furthermore, we identified optical \ion{He}{i} singlet emission lines
   emitted in the broad-line region.
   The observed line fluxes agree with theoretical predictions.
   We show that a broad wing on the red side of the   
   [\ion{O}{iii}]$\lambda$5007 line is caused by the \ion{He}{i} singlet
   line at 5016 \AA.
   }
   
  \authorrunning{W. Kollatschny et al.}
  \titlerunning{Short-term variations in Mrk\,110}
   \maketitle
%
%________________________________________________________________

\section{Introduction}

   The variability of the continuum and broad emission lines 
in Seyfert 1 galaxies was detected more than 20 years ago.
   The delay of the line intensity variations with respect to the
   varying ionizing continuum yields information on the extent and
   structure of the innermost broad-line region (BLR) in AGN.
   During the past 10 to 15 years international collaborations 
   or individual groups studied the optical variations of
   more than half a dozen of
   Seyfert galaxies such as e.g. NGC\,5548 (International AGN Watch -- Peterson et al. \cite{peterson94}),
   NGC4151 (Wise observatory group -- Maoz et al. \cite{maoz91}) and
   NGC\,4593 (LAG collaboration -- Robinson \cite{robinson94};
   Kollatschny \& Dietrich \cite{kollatschny97}).
   The main difficulties encountered by these campaigns have been the 
   inhomogeneity of the observed spectra and the 
   required high S/N of the data. Furthermore the campaigns must extend
   over time-scales of many months to years with dense
  temporal sampling of days to weeks. 

   In this paper we discuss high S/N spectra of the narrow-line Seyfert galaxy
   Mrk\,110 resulting from
   a variability campaign with the 9.2-m Hobby-Eberly Telescope.
   It has been shown before that this galaxy shows extreme 
   variability amplitudes (Peterson et al.\ \cite{peterson98},
    Bischoff \& Kollatschny
    \cite{bischoff99}) on time scales of weeks to months.
   This new campaign improves upon the past studies of this galaxy
   with respect to homogeneity of the data, S/N ratio of the spectra, 
   temporal sampling and coverage of the optical spectral range.
   Furthermore, the present data allow the determination    
   of the central black hole mass with high precision
  by using the variations of many different emission lines.        

  The strongest forbidden lines in optical AGN spectra are the
  lines of [\ion{O}{iii}]$\lambda$4959,5007.
  Meyers \& Peterson (\cite{meyers85}), van Groningen \& de Bruyn
  (\cite{groningen89}), Osterbrock (\cite{osterbrock85}), and
  others have detected a broad feature on the red wing of the
  [\ion{O}{iii}]$\lambda$5007 line in a significant number of Seyfert
  galaxies. Meyers \& Peterson (\cite{meyers85}) and van Groningen \& de Bruyn
  (\cite{groningen89})
  ruled out the possibility that the wing was caused by broad
  \ion{Fe}{ii} blends or the \ion{He}{i}$\lambda$5016 singlet line,
 and designated
  the wing as [\ion{O}{iii}] from the broad-line region (BLR) in the AGN.
  A clear confirmation of this identification would have important
  consequences for the determination of gas density in the BLR.
  A search for an identical broad wing on the red side 
  of [\ion{O}{iii}]$\lambda$4959 in high S/N spectra would
 test this idea, but to date no such wing has been seen.
%
%__________________________________________________________________

\section{Observations and data reduction}

 We obtained 26 spectra  of Mrk\,110 with the 9.2-m Hobby-Eberly
 Telescope (HET) at McDonald Observatory % as part of the science queue.
 between 1999 November 13 and 2000 May 14.
 Table 1 lists the observing dates.
\begin{table}
\caption{Log of observations}
\begin{tabular}{ccc}
\hline 
\noalign{\smallskip}
Julian Date & UT Date & Exp. time \\
2\,400\,000+&         &  [sec.]   \\
%(1) & (2) & (3) & (4) \\ %& (5) & (6) & (7) & (8) & (9) \\ 
\noalign{\smallskip}
\hline 
\noalign{\smallskip}
51495.94 & 1999-11-13 & 1200 \\
51497.91 & 1999-11-15 & 1200 \\
51500.91 & 1999-11-18 & 1200 \\
51518.89 & 1999-12-06 & 1200 \\
51520.87 & 1999-12-08 & 1200 \\
51522.88 & 1999-12-10 & 1200 \\
51525.84 & 1999-12-13 & 1200 \\
51528.84 & 1999-12-16 & 1200 \\
51547.80 & 2000-01-04 & 1200 \\
51584.72 & 2000-02-10 & 1200 \\
51586.71 & 2000-02-12 & 1200 \\
51595.88 & 2000-02-21 & 1200 \\
51598.86 & 2000-02-24 & 1200 \\
51605.83 & 2000-03-02 & 1200 \\
51608.62 & 2000-03-05 & 1200 \\
51611.62 & 2000-03-08 & 1200 \\
51614.63 & 2000-03-11 & 1200 \\
51629.76 & 2000-03-26 &  540 \\
51637.77 & 2000-04-03 &  600 \\
51645.73 & 2000-04-11 &  600 \\
51658.70 & 2000-04-24 &  600 \\
51663.68 & 2000-04-29 &  540 \\
51664.66 & 2000-04-30 &  600 \\
51670.70 & 2000-05-06 &  360 \\
51673.69 & 2000-05-09 &  570 \\
51678.64 & 2000-05-14 &  600 \\
\noalign{\smallskip}
\hline 
\end{tabular}
\end{table}
The 26 spectra were obtained over a period of 182.7 days. The average interval between the observations was 7.3 days and the median interval was only 3.0 days.

 All observations were made under identical instrumental conditions with the
 Marcario Low Resolution Spectrograph
 (LRS) (Hill et al. \cite{hill98}, Cobos et al. \cite{cobos98})
 located at the prime focus. We used
 a Ford Aerospace 3072x1024 CCD with 15 $\mu$m pixel in 2x2 binning. 
 The slit width was fixed to 2\arcsec\hspace*{-1ex}.\hspace*{0.3ex}0 projected on the sky, and the position angle of the slit was set to PA=45\degr\ throughout the 
 campaign. The spatial resolution corresponds to
 0.472\arcsec per binned pixel. We extracted our object spectra over 7 pixels, 
 corresponding to 3.3\arcsec.
 The resolving power %of our LRS grism 2 configuration
 was 650 and the spectra cover the wavelength range from 4200\,\AA\
 to 6900~\AA\ in the rest frame of the galaxy.
 The majority of the  observations were comprised of two
 10 minute integrations, which in most cases yielded a     
 S/N  $>$  100 per pixel in the continuum.

 HgCdZn and Ne spectra were taken after each object exposure for
 wavelength calibration. Spectra of different standard stars were
 observed for flux calibration.
 The reduction of the spectra (bias subtraction, cosmic ray correction,
 flat-field correction, 2D-wavelength calibration, night sky subtraction,
 flux calibration etc.) was done in a homogeneous way with IRAF reduction
 packages. 
 The spectra have not been corrected for atmospheric absorption in the B band.

 Great care was taken to produce good intensity and wavelength calibrations.
 All spectra were calibrated to the same absolute
 [\ion{O}{iii}]$\lambda$5007 flux of 2.26 $10^{-13}$ \ergscm\
 (Bischoff \& Kollatschny \cite{bischoff99}, Peterson et al.\ \cite{peterson98}).
 The spatially unresolved structure of the narrow-line region in Mrk\,110
 has been verified before (e.g. Bischoff \& Kollatschny \cite{bischoff99})
 for utilizing this internal calibration method. 
 The accuracy  of the [\ion{O}{iii}]$\lambda$5007 flux calibration
 was tested on all forbidden emission lines in the spectra.
 We calculated difference spectra of all epochs
 with respect to the mean spectrum of our variability campaign.
 Corrections for small spectral shifts ($<$ 0.5 \AA )
 and for small scaling factors were executed
 by minimizing the residuals of the narrow emission lines in the
 difference spectra. 
 All wavelengths were converted to the rest frame of the galaxy (z=0.0355).  
 We obtained a relative flux  accuracy of better than 1\% in  most of
 our spectra.

\section{Results and discussion}

\subsection{Line and continuum variations}

\begin{figure*}
 \hbox{\includegraphics[bb=40 90 380 700,width=9.12cm,angle=270]{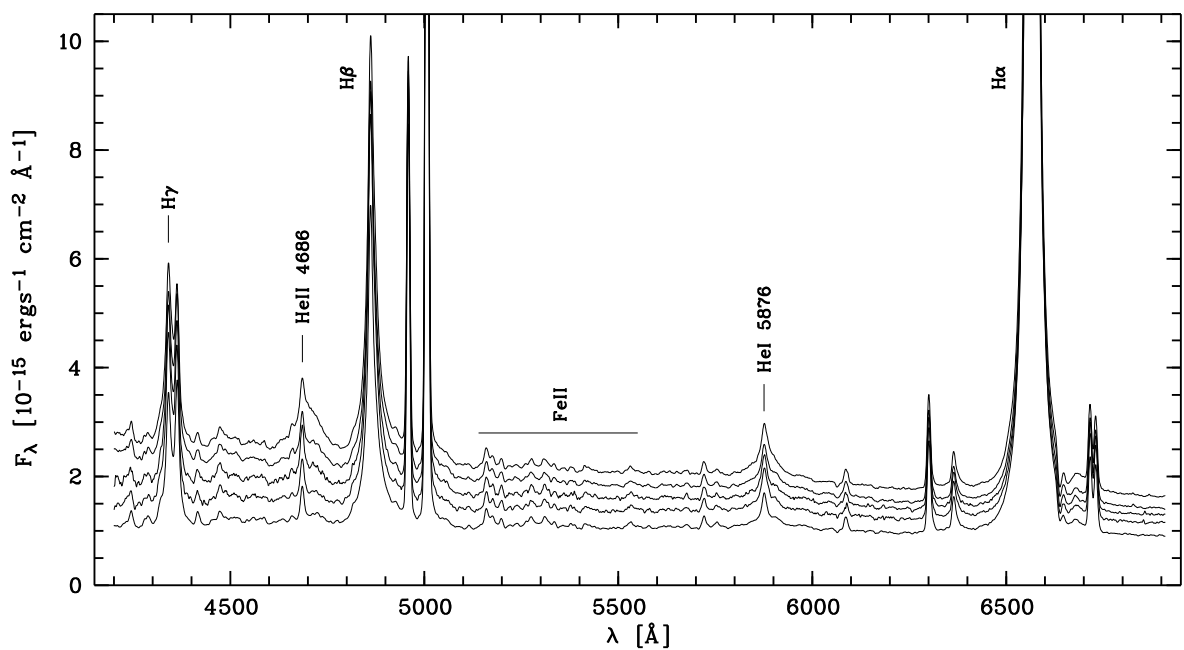}}
       \vspace*{-2mm} 
  \caption{HET spectra of Mrk\,110 taken at 2000 Jan. 4,
   1999 Dec. 16, 2000 Feb. 24, 2000 March 8, 2000 April 4 (from top to
  bottom).}
   \label{m110spekvar}
\end{figure*}

 Rest frame spectra of Mrk\,110 obtained during our new campaign are shown in 
 Fig.~\ref{m110spekvar}. The strongest broad emission lines are labeled.
%
%------------------------------------------------------------------------------
%
%
%----------------------------------------------------------------------------- 
%  
%------------------------------------------------------------------------------
%
\begin{figure*}
 \hbox{\includegraphics[bb=40 90 380 700,width=9.12cm,angle=270]{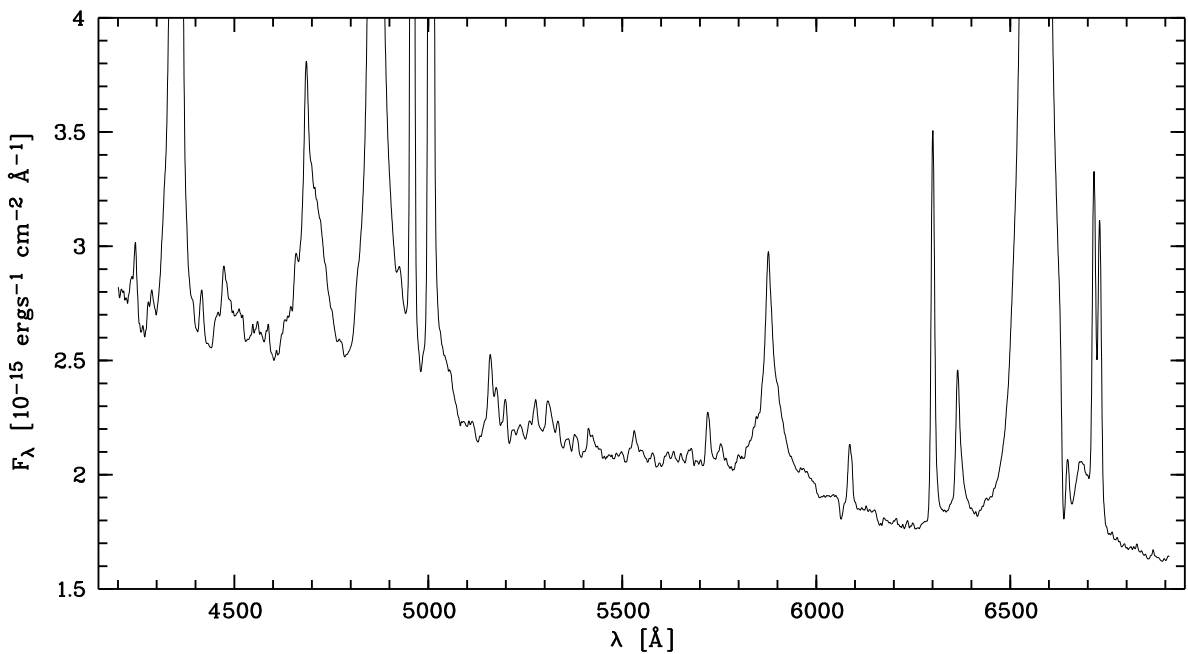}}
       \vspace*{-2mm} 
  \caption{An individual HET spectrum taken 2000 Jan.4 emphasizing
       the weak lines.}
   \label{m110spekvarblowup}
\end{figure*}
%
%----------------------------------------------------------------------------- 
%  
 The continuum variations, the variations
 of the continuum gradient, and the variations of the broad
 \ion{He}{ii} $\lambda4686$ line are obvious. On the other hand the  
 \ion{Fe}{ii} multiplet lines remained constant.
 In Fig.~\ref{m110spekvarblowup} an individual spectrum is plotted
 emphasizing the weak lines.

The wavelength boundaries we used for our continuum measurements
are given in Table 2.
\setcounter{table}{1}
\begin{table}
\caption{Boundaries of mean continuum values and line integration limits
(H$\gamma$ incl.[\ion{O}{iii}], H$\alpha$ incl.[\ion{N}{ii}])}

\begin{tabular}{lccccccc}
\hline 
\noalign{\smallskip}
Cont./Line & Wavelength range & Pseudo-continuum \\
\noalign{\smallskip}
(1) & (2) & (3) \\
\noalign{\smallskip}
\hline 
\noalign{\smallskip}
Cont.~4265         & 4260\,\AA\ -- 4270\,\AA \\
Cont.~5135         & 5130\,\AA\ -- 5140\,\AA \\
Cont.~6895         & 6890\,\AA\ -- 6900\,\AA \\
H$\gamma$          & 4300\,\AA\ -- 4400\,\AA & 4270\,\AA\ -- 4430\,\AA \\
HeI$\lambda 4471$  & 4435\,\AA\ -- 4535\,\AA & 4435\,\AA\ -- 4535\,\AA \\
HeII$\lambda 4686$ & 4600\,\AA\ -- 4790\,\AA & 4600\,\AA\ -- 5130\,\AA \\
H$\beta$           & 4790\,\AA\ -- 4940\,\AA & 4600\,\AA\ -- 5130\,\AA \\
HeI$\lambda 5016$  & 4980\,\AA\ -- 5090\,\AA & 4600\,\AA\ -- 5130\,\AA \\
HeI$\lambda 5876$  & 5785\,\AA\ -- 6025\,\AA & 5650\,\AA\ -- 6120\,\AA \\
H$\alpha$          & 6420\,\AA\ -- 6700\,\AA & 6260\,\AA\ -- 6780\,\AA \\
\noalign{\smallskip}
\hline \\
\end{tabular}
\end{table}
These regions are free of strong emission and/or absorption lines. 
The broad emission line intensities were integrated between the listed limits
after subtraction of a linear pseudo-continuum defined by the boundaries
given in Column\,(3).

The results of the continuum and line intensity measurements are given in 
Table\,3. We
determined the line intensities with respect to the mean spectrum of Mrk\,110
(Fig.~\ref{het1mean}).
Thus the measurement of the pseudo-continuum is not hurt
by the presence of any underlying
 absorption lines from the 
host galaxy or any constant emission lines.
But one has to keep in mind that the line fluxes given in Table\,3
still contain the narrow-line components within the listed
wavelength boundaries. Thus the true fractional variability amplitudes
of the broad lines are larger than those listed in Table\,4.
The errors given in Table\,3 are the internal errors of our measurements 
with respect to the mean spectrum.
A few spectra are of lower quality because of poorer transparency due to
clouds.

\begin{figure*}
 \hbox{\includegraphics[bb=40 90 380 700,width=9.12cm,angle=270]{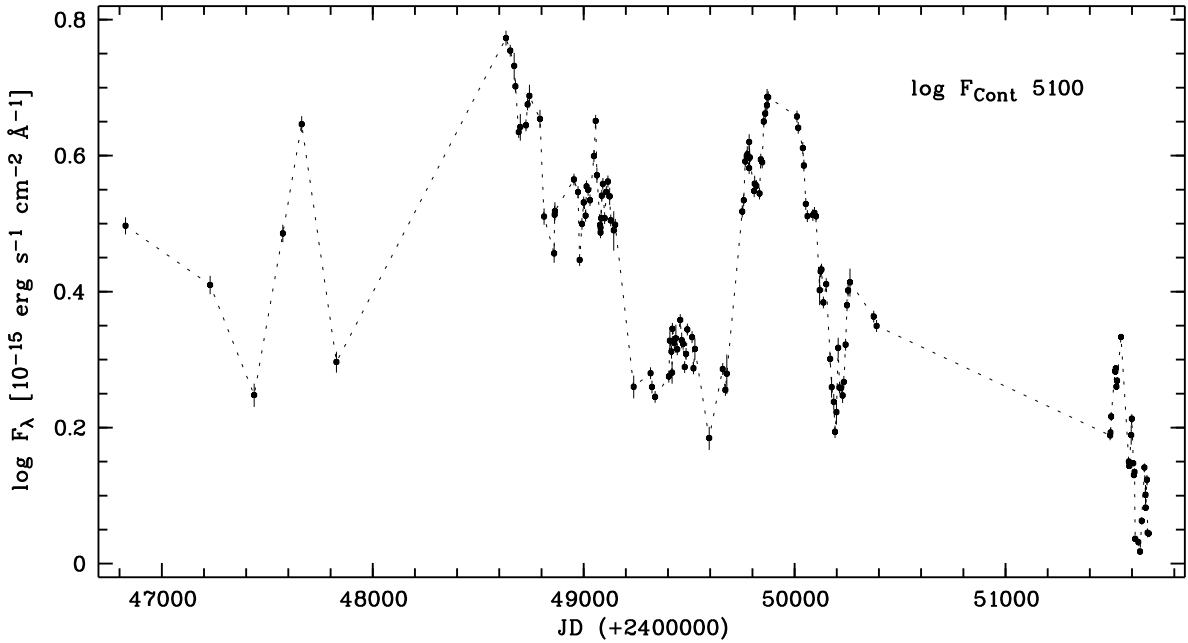}}
       \vspace*{-2mm} 
  \caption{Long-term continuum light curve at 5100\,\AA\
 from 1987 until 2000 May. The points are connected by a dotted line to aid
 the eye. }
   \label{longtermlc}
\end{figure*}

The light curves of the continuum flux at 5135\,\AA\
and of the integrated Balmer and Helium emission line intensities
are shown in Fig.\,3.
%
%------------------------------------------------------------------------------
%
\begin{figure*}
 \hbox{\includegraphics[bb=40 90 380 700,width=55mm,height=85mm,angle=270]{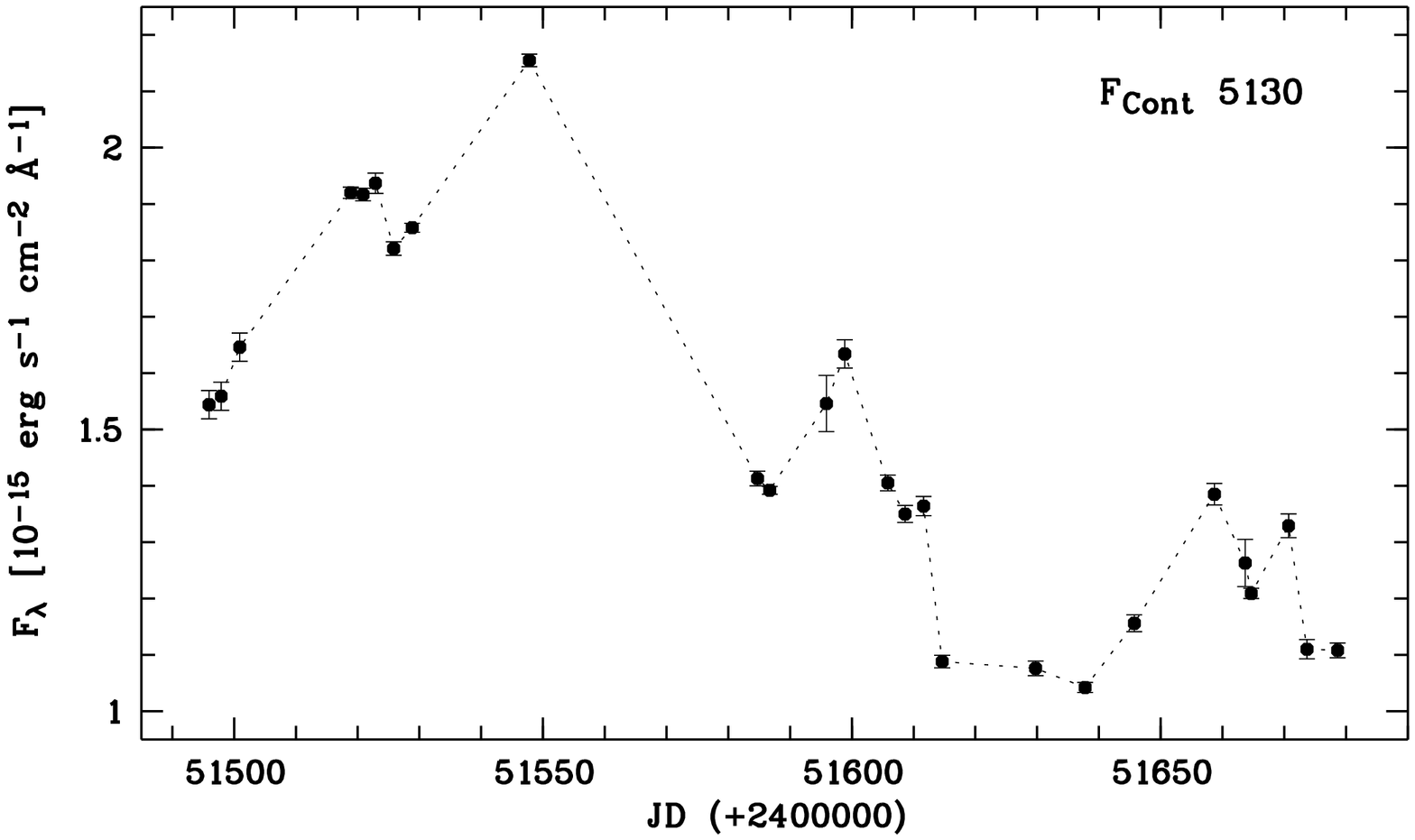}\hspace*{7mm}
       \includegraphics[bb=40 90 380 700,width=55mm,height=85mm,angle=270]{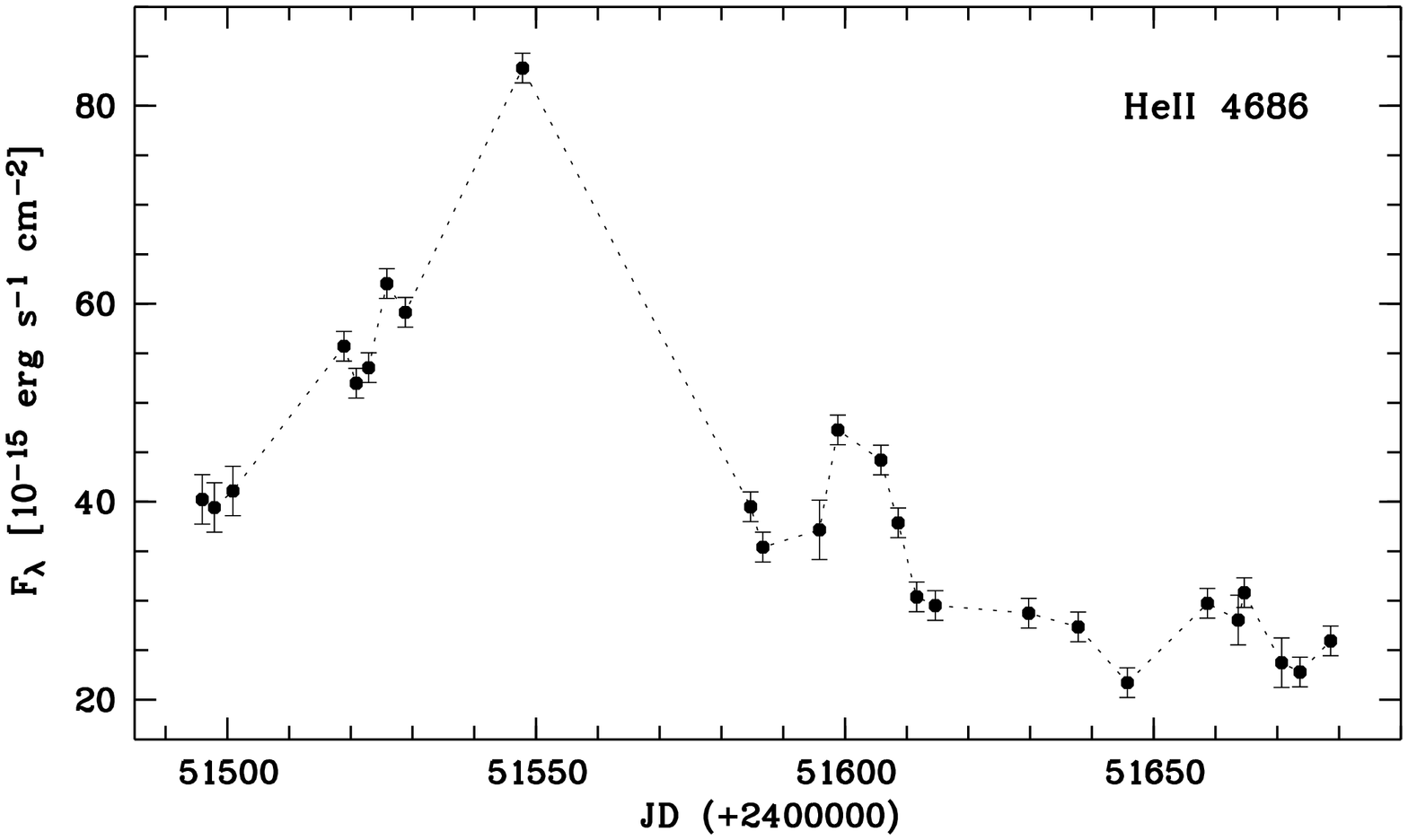}}
 \hbox{\includegraphics[bb=40 90 380 700,width=55mm,height=85mm,angle=270]{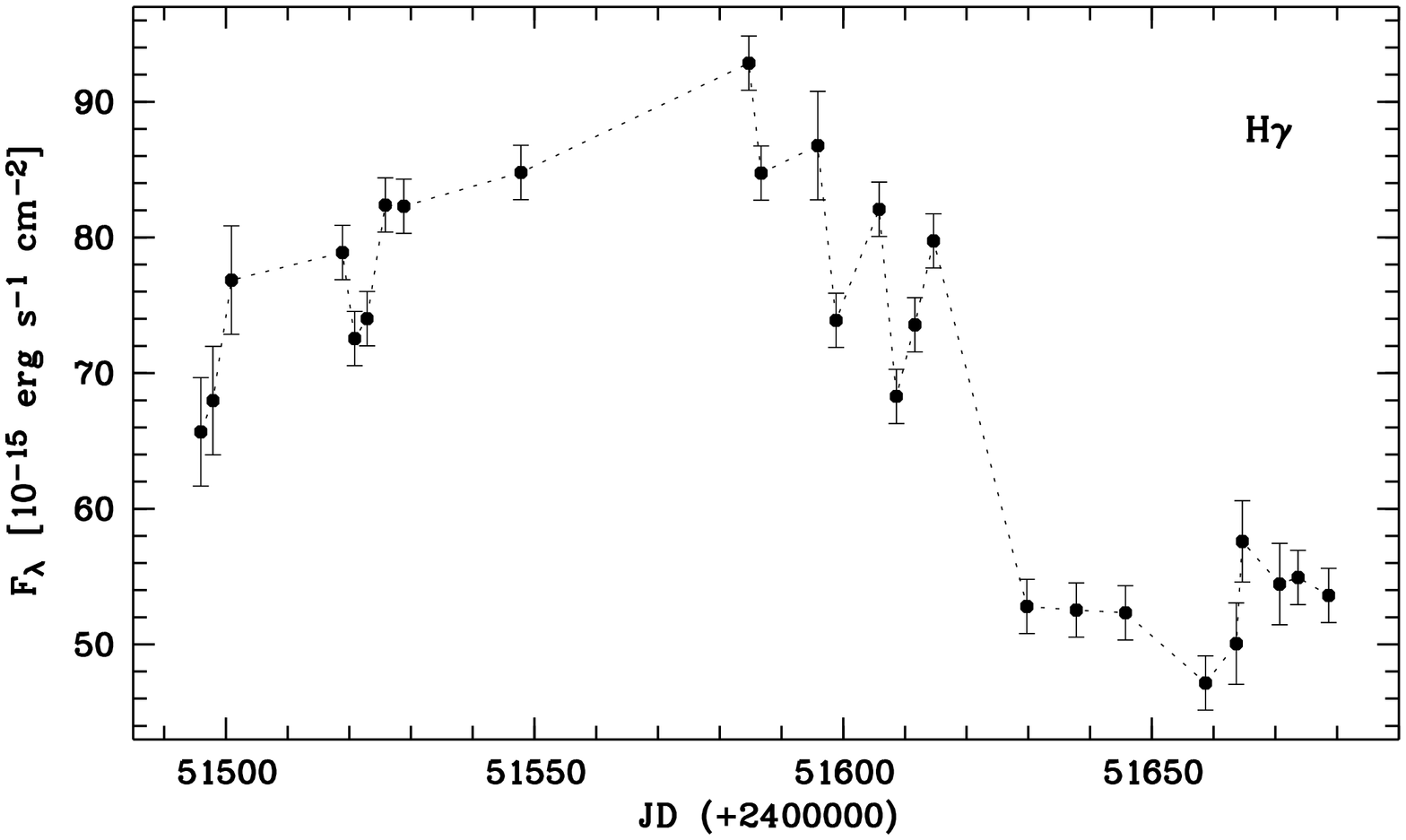}\hspace*{7mm}
       \includegraphics[bb=40 90 380 700,width=55mm,height=85mm,angle=270]{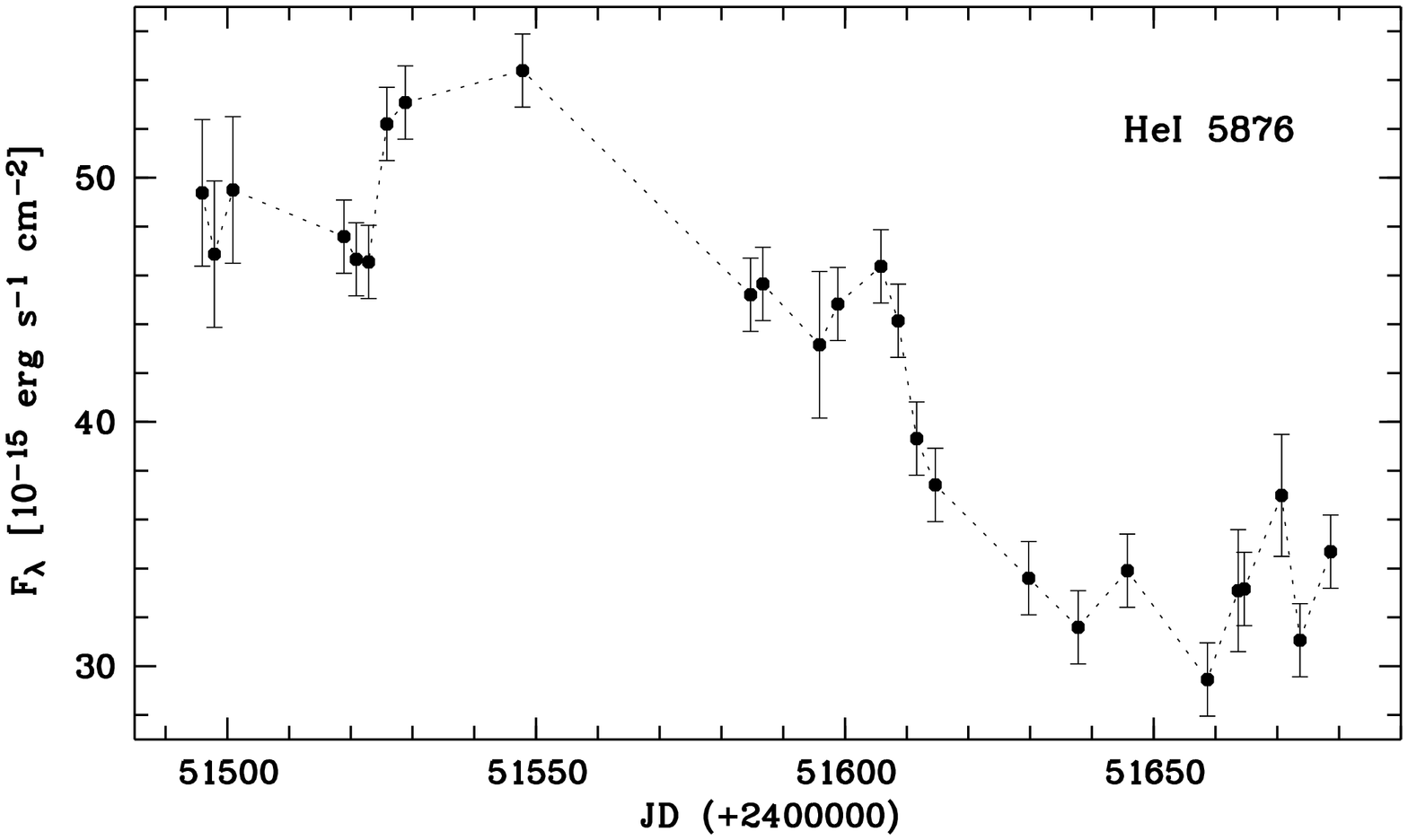}}
 \hbox{\includegraphics[bb=40 90 380 700,width=55mm,height=85mm,angle=270]{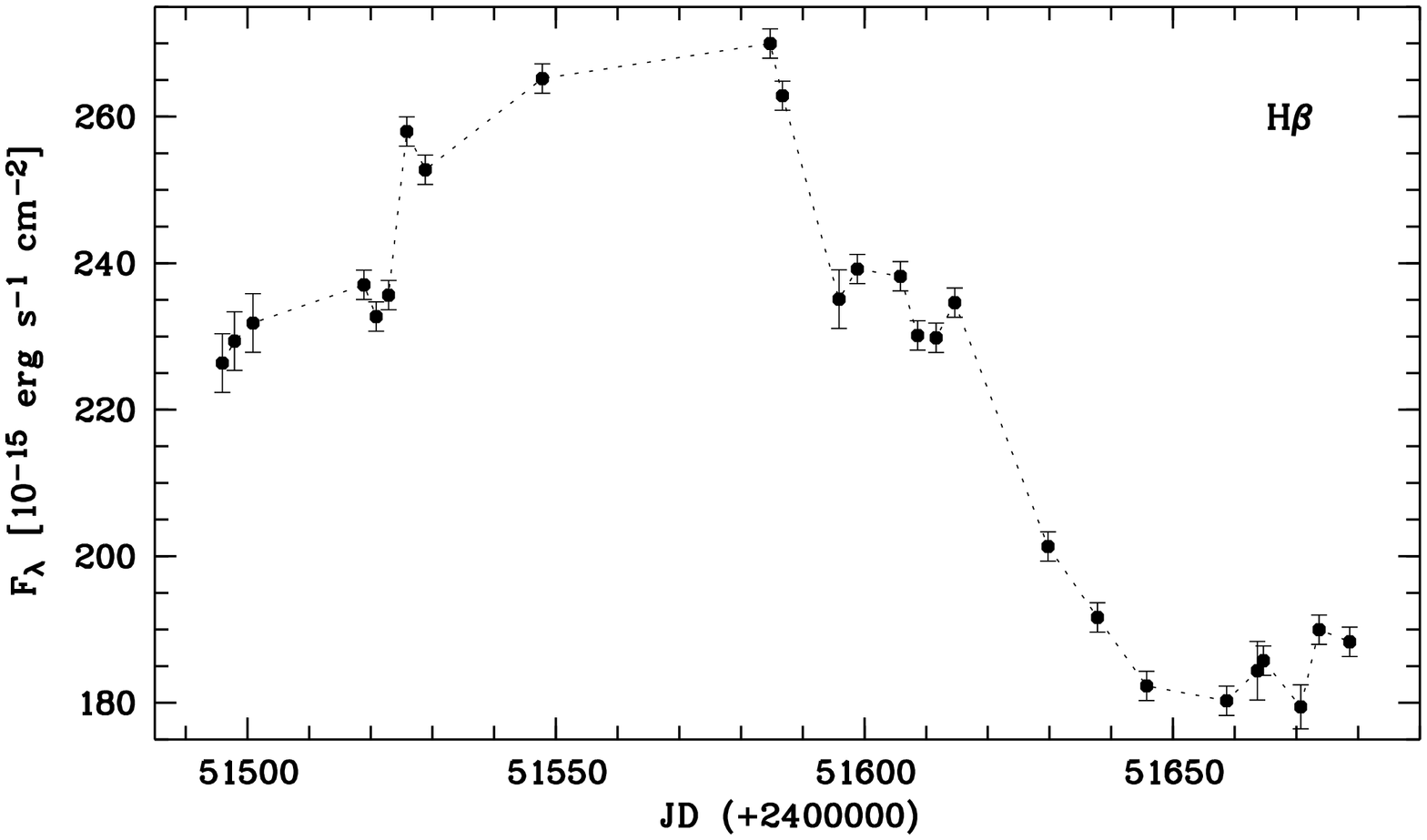}\hspace*{7mm}
       \includegraphics[bb=40 90 380 700,width=55mm,height=85mm,angle=270]{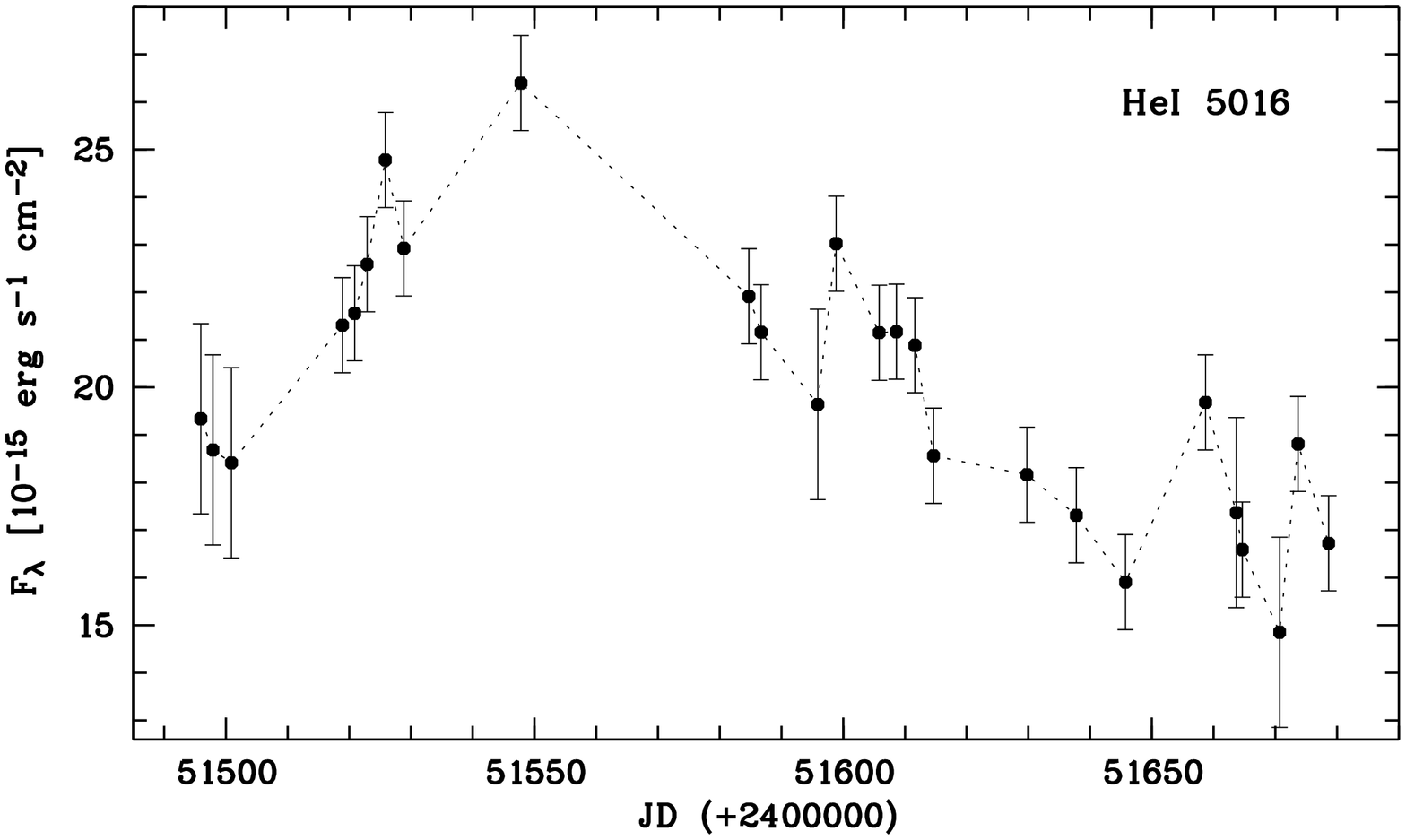}}
 \hbox{\includegraphics[bb=40 90 380 700,width=55mm,height=85mm,angle=270]{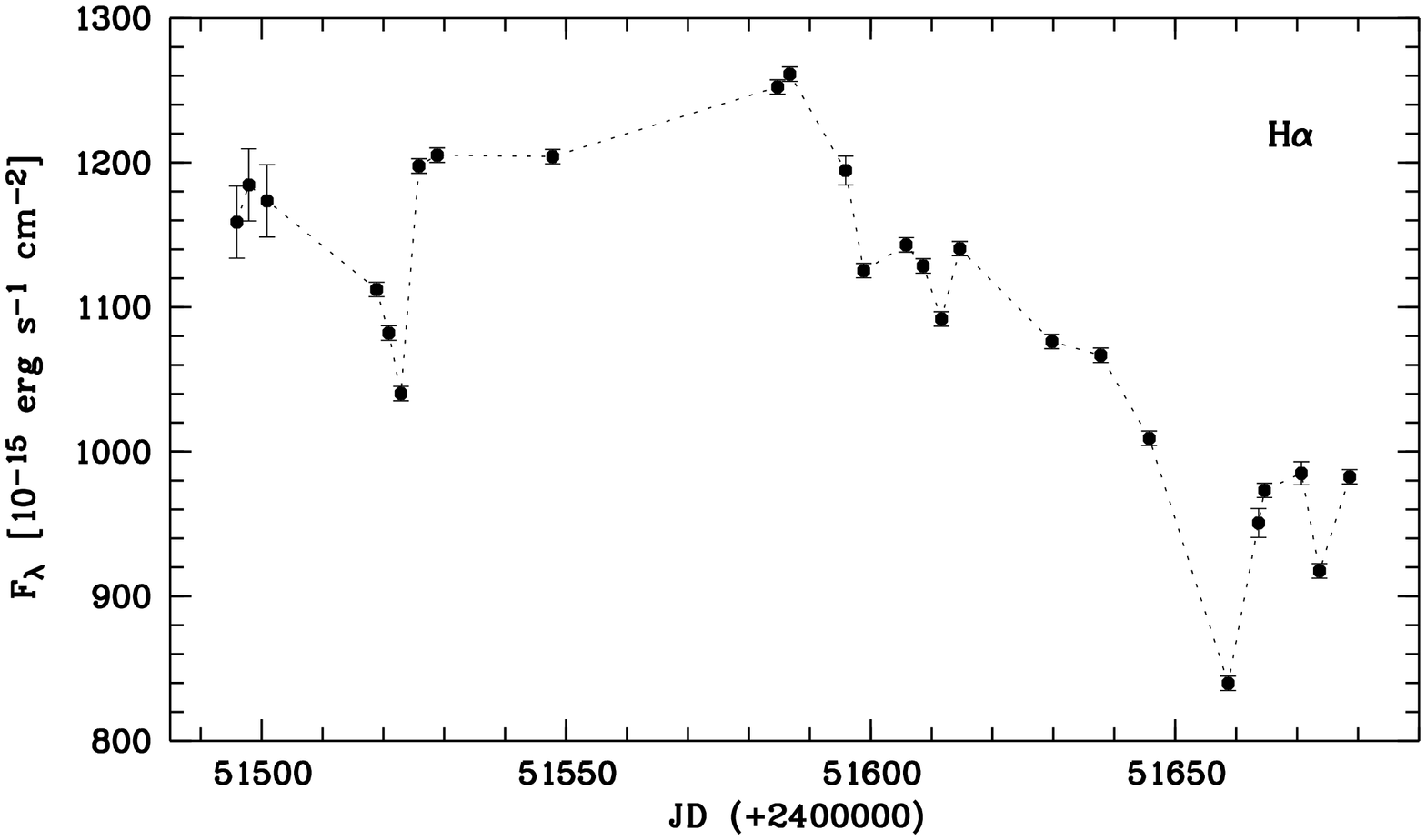}\hspace*{7mm}
       \includegraphics[bb=40 90 380 700,width=55mm,height=85mm,angle=270]{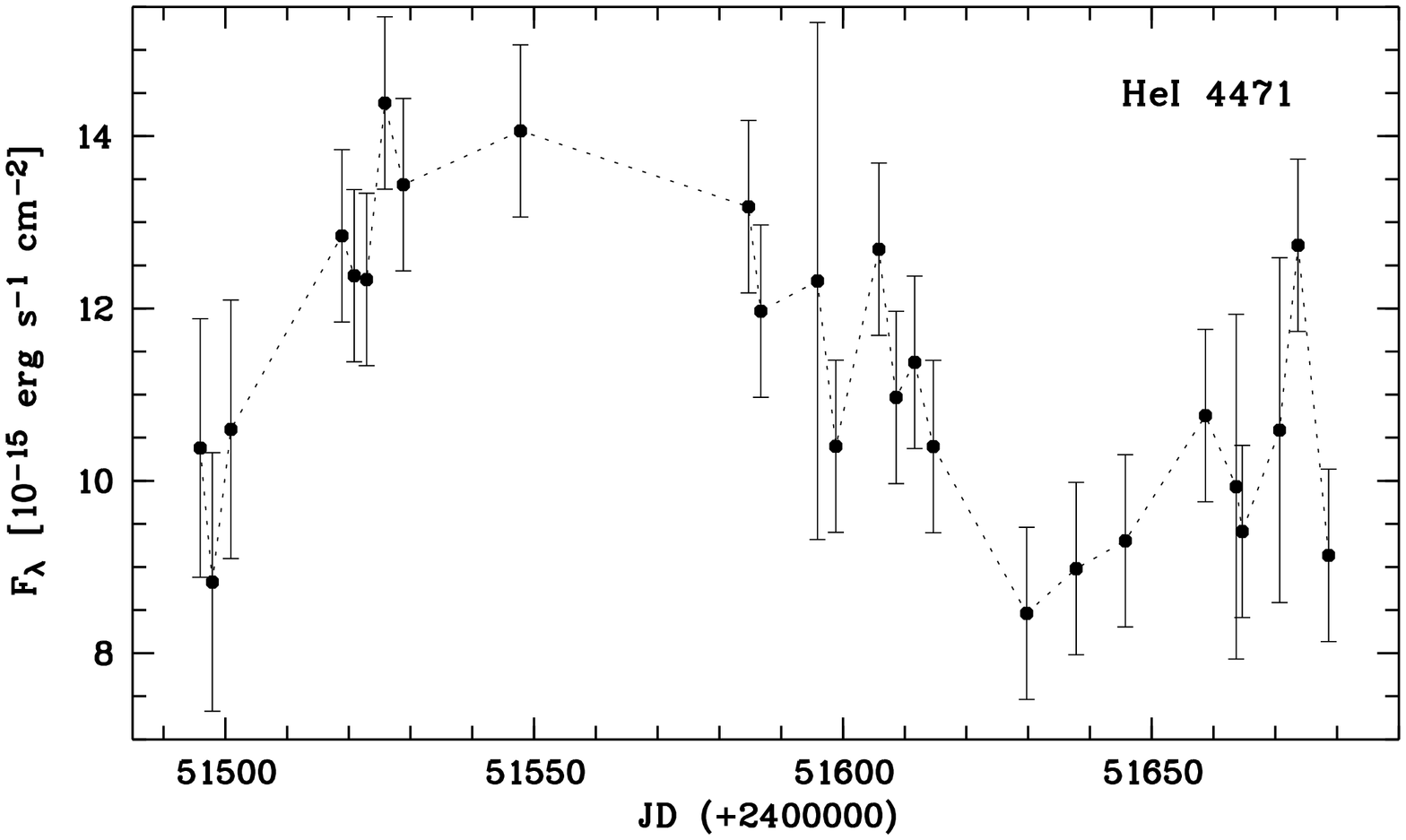}}
       \vspace*{5mm} 
  \caption{Light curves of continuum flux at 5135\,\AA\
 (in units of 10$^{-15}$ erg cm$^{-2}$ s$^{-1}$\,\AA$^{-1}$) and of integrated
  emission line fluxes of H$\gamma$  H$\beta$, H$\alpha$,
  HeII$\lambda$4686, HeI$\lambda$5876, HeI$\lambda$5016, and
 HeI$\lambda$4471 (in units of  10$^{-15}$ erg cm$^{-2}$ s$^{-1}$). The points 
 are connected by a dotted line to aid the eye.}
\end{figure*}
%
%------------------------------------------------------------------------------
%
 Fig.~\ref{longtermlc} shows the logarithm of the long-term continuum
 variations of Mrk\,110 from 1987 February until 2000 May. The data are
 taken from Bischoff \& Kollatschny (\cite{bischoff99}), Peterson et al.\
 (\cite{peterson98}), this paper and from a newly reduced
 data point at JD 2\,448\,793. 
 Besides the long-term behaviour of the continuum and the strong variations
 at certain epochs  one can recognize immediately that 
 Mrk\,110 was in a very low state during
 the HET variability campaign (after JD 2\,451\,500). In 2000
 March/April the continuum was at a historical low.
%
%------------------------------------------------------------------------------
%
%
%------------------------------------------------------------------------------
%
\setcounter{table}{3}
\begin{table}
\caption{Variability statistics for Mrk\,110 in 1999/2000}
\begin{tabular}{lccccccc}
\hline 
\noalign{\smallskip}
Cont./Line & F$_{min}$ & F$_{max}$ & R$_{max}$ & $<$F$>$ & $\sigma_F$ & F$_{var
}$ \\
\noalign{\smallskip}
(1) & (2) & (3) & (4) & (5) & (6) & (7) \\ %(8) & (9) \\ 
\noalign{\smallskip}
\hline 
\noalign{\smallskip}
Cont.~4265         & 1.07 &  2.63  & 2.46 & 1.66  & 0.451 & 0.270 \\
Cont.~5135         & 1.04 &  2.16  & 2.07 & 1.47  & 0.314 & 0.213 \\
Cont.~6895         & 0.91 &  1.63  & 1.79 & 1.18  & 0.193 & 0.163 \\
HeII$\lambda 4686$ & 21.7 & 83.8   & 3.86 & 39.5  & 14.6  & 0.368 \\
HeI$\lambda 4471$  & 8.46 & 14.4   & 1.70 & 11.2  & 1.70  & 0.099 \\
HeI$\lambda 5016$  & 14.9 & 26.4   & 1.78 & 20.0  & 2.77  & 0.123 \\
HeI$\lambda 5876$  & 29.5 & 54.4   & 1.85 & 41.9  & 7.56  & 0.175 \\
H$\gamma$          & 47.2 & 92.9   & 1.97 & 69.3  & 13.7  & 0.194 \\
H$\beta$           & 179. & 270.   & 1.51 & 223.  & 29.1  & 0.130 \\
H$\alpha$          & 840. & 1261.  & 1.50 & 1096. & 108.  & 0.098 \\
\noalign{\smallskip}
\hline 
\end{tabular}

Continuum flux in units of
 10$^{-15}$\,erg\,sec$^{-1}$\,cm$^{-2}$\,\AA$^{-1}$.\\
Line flux in units of 10$^{-15}$\,erg\,sec$^{-1}$\,cm$^{-2}.$
\end{table}

In Table 4 we list
statistics of the continuum and emission line variations.
Given are the minimum and maximum fluxes F$_{min}$ and F$_{max}$,
peak-to-peak amplitudes R$_{max}$ = F$_{max}$/F$_{min}$, the mean flux
over the period of observations $<$F$>$, the standard deviation $\sigma_F$,
 and the fractional variation 
\[ F_{var} = \frac{\sqrt{{\sigma_F}^2 - \Delta^2}}{<F>} \] 
as defined by Rodr\'iguez-Pascual et al.\ (\cite{rodriguez97}).

The variability amplitude of the continuum flux in this campaign
was about one half of the variations
over the past ten years (see Fig.~\ref{longtermlc}). The same 
trend is seen in the variability amplitudes of all emission line fluxes.
The extreme variability amplitudes of Mrk~110 
compared to other galaxies have been mentioned before
(Bischoff \& Kollatschny, \cite{bischoff99}).
The variability amplitude of the blue continuum flux is 
stronger than that of the red continuum,
both in a relative and in an absolute flux sense 
(Table\,4, Columns\,4,7).

\subsection{Mean and rms spectra}

\begin{figure*}
    \includegraphics[bb=40 90 380 700,width=9.12cm,angle=-90]{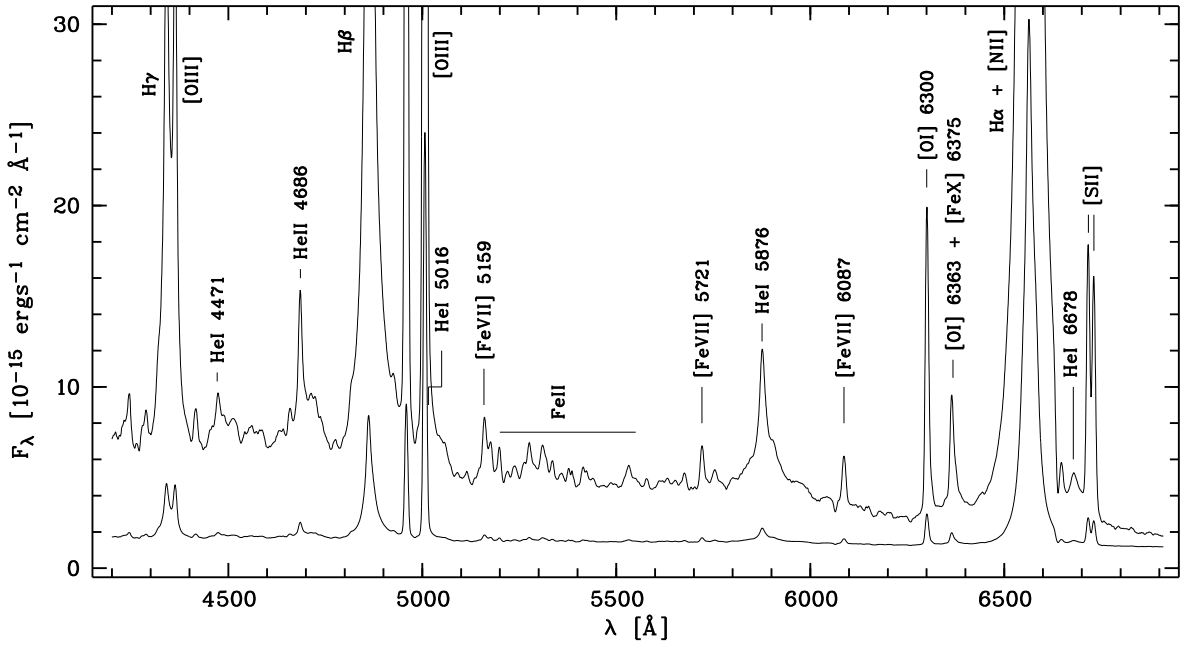}
      \caption{Mean rest frame spectrum of Mrk\,110
               for 24 epochs from Nov. 1999 through May 2000. 
               The upper spectrum is scaled by a factor
               of 10 (zero level is shifted by -10) to show both strong
               and weak lines.
              }
         \label{het1mean}
   \end{figure*}

 The mean rest frame spectrum of Mrk\,110 is shown in Fig.~\ref{het1mean}
 with two different vertical scalings to show both strong and weak lines.
%                                                
%----------------------------------------------------------- mean
%
%______________________________________________________________
%
 This spectrum has been derived from spectra obtained at 24 epochs.
 Two of the 26 spectra (2000 February 21 and April 30)
 are not included in the average because of a lower S/N and poor
 wavelength calibration, respectively.
 The Balmer and some He emission lines
 are labeled. One can  clearly identify individual \ion{Fe}{ii} lines
 in the \ion{Fe}{ii}$\lambda$5200 blend.

\begin{figure}
    \includegraphics[bb=40 60 570 780,width=6.5cm,angle=-90]{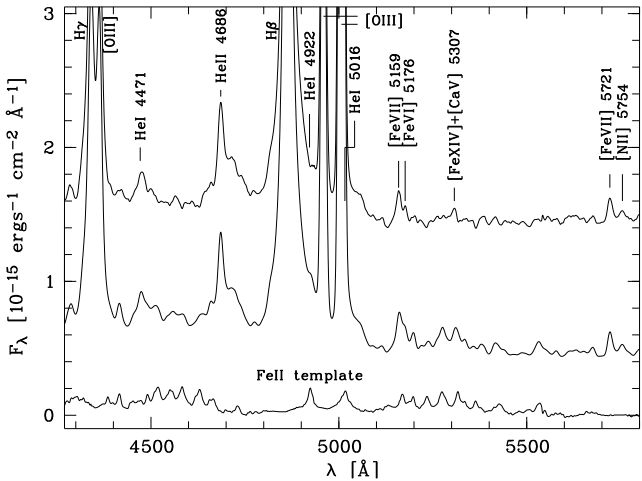}
      \caption{Blue spectral range of Mrk\,110: original mean spectrum
               vertically shifted by -1 (middle), scaled \ion{Fe}{ii}
               template spectrum (bottom), and
               \ion{Fe}{ii} subtracted spectrum (top).
              }
         \label{het1fe}
   \end{figure}

 The mean spectrum contains
 heavily blended \ion{Fe}{ii} lines. 
 We subtracted a scaled \ion{Fe}{ii} template spectrum 
 to estimate the contribution of these lines
 and to identify further weak lines. 
 This \ion{Fe}{ii} template spectrum has been 
 derived from the PG quasar sample (Boroson \& Green \cite{boroson92}).
 We used the \ion{Fe}{ii}$\lambda$5198, $\lambda$5276 and
 $\lambda$5535 lines of multiplets 49, 48 and 55 to scale the template 
 spectrum %and to look for residuals in the difference spectrum.
 and broadened our Mrk\,110 spectrum slightly with a Gaussian. 
 In Fig.~\ref{het1fe} our broadened Mrk\,110 spectrum is shown in the middle.
 The scaled \ion{Fe}{ii} template spectrum is shown at the bottom.
 The \ion{Fe}{ii} subtracted spectrum is plotted at the top.
%
%----------------------------------------------------------- fe
%
%______________________________________________________________
%
 Some highly ionized [\ion{Fe}{vi}], [\ion{Fe}{vii}] and [\ion{Fe}{xiv}] lines
 clearly stick out in the \ion{Fe}{ii} subtracted spectrum.
 The line strengths are similar to their strengths in  IIIZw\,77. This
 AGN has been classified by Osterbrock (\cite{osterbrock81}) 
 as an unusual, high-ionization Seyfert 1 galaxy.
 The \ion{He}{i}$\lambda4471$ line is also clearly revealed once the 
 underlying \ion{Fe}{ii} emission is subtracted. Notice the broad wing
 on the red side of the narrow [\ion{O}{iii}]$\lambda$5007 line remains
 nearly  the same before and after \ion{Fe}{ii} subtraction.
 Therefore, this line feature cannot be attributed
 to \ion{Fe}{ii} line blends.  

\begin{figure*}
    \includegraphics[bb=40 90 450 700,width=11cm,clip,angle=-90]{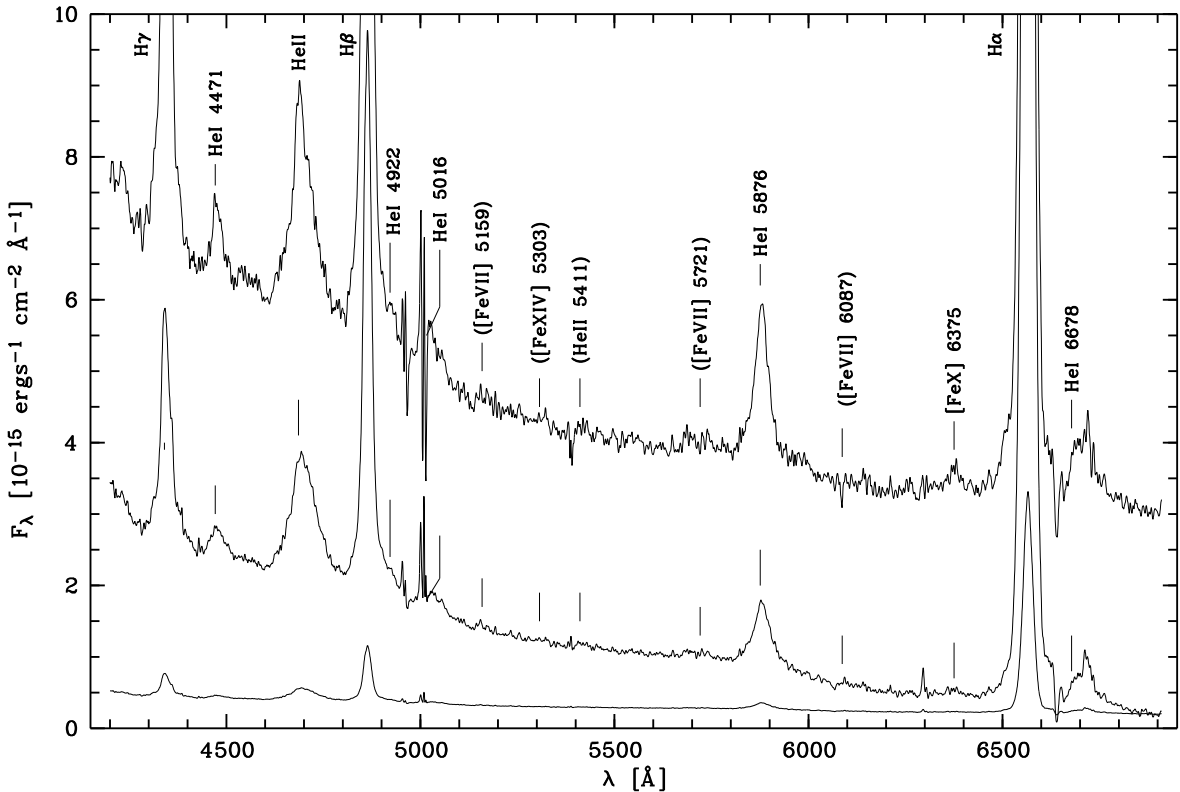}
      \caption{Rms spectrum of Mrk\,110 (bottom). The rms spectrum is 
              plotted with a vertical magnification of 10 (zero level   
              is shifted by -0.2) to show both strong and weak lines
              in the middle. A difference spectrum (mean maximum stage
              minus mean minimum stage as described in the text) is shown 
               on the top for comparison.
              For completeness marginally detected lines are given in
              parentheses.}
         \label{het1rms}
   \end{figure*}

 Fig.~\ref{het1rms}
 shows the rms spectrum of Mrk\,110 derived from the same 24 epochs
 as the mean spectrum, with two
 different scalings.
%
%----------------------------------------------------------- rms
%
%______________________________________________________________
%
 The rms spectrum enhances the variable parts of the spectrum.
 Lines with constant flux and profile,
 such as the [\ion{O}{iii}] lines,
 disappear in the rms spectrum.
 But the rms spectrum
 amplifies noise as well as small inaccuracies in the wavelength
 and flux calibration, producing sharp, high frequency artifacts.
 This can be seen in the very
 strong [\ion{O}{iii}]$\lambda\lambda$5007,4959 lines.

 The rms spectrum clearly shows the
 H$\alpha$, H$\beta$, and  H$\gamma$ Balmer lines,
 the broad \ion{He}{ii}$\lambda4686$
 line and  the \ion{He}{i} lines at
 $\lambda4471$, $\lambda4922$, $\lambda5016$,
 $\lambda5876$ and $\lambda6678$.
 The flux of the 
 \ion{He}{i}$\lambda4922$ and  \ion{He}{i}$\lambda6678$ lines
 can be determined by subtracting the
 blue side of the Balmer profiles from the red one 
 after flipping the profile around their central wavelengths.
 The \ion{He}{i}$\lambda6678$ line
 is heavily blended by atmospheric B-band absorption
 and therefore must be treated with caution.
 But the proof of identity has been shown
 in the mean spectrum independently.
 
 In Fig.~\ref{het1rms} (top) we also
 show the difference between the mean high stage spectrum and the
 mean low stage spectrum
 as an additional test of line variability.
 The mean high stage spectrum has been deduced from all spectra
 obtained from 1999 Dec. through 2000 March 11, and
 the mean low stage spectrum from all spectra
 from 2000 March 26 through the end of the campaign in 2000 May.
 Again we did not consider the lower quality spectra taken
 at 2000 Feb. 21 and April 3. 
 One can unambiguously identify the same lines as in 
 the rms spectrum including the variable [\ion{Fe}{x}]$\lambda6375$ line.
 Furthermore,  in the difference spectra
 of our long-term variability campaign it was to be seen
 (Bischoff \& Kollatschny \cite{bischoff99}, Fig.\,3)
 that the [\ion{Fe}{x}]$\lambda6375$ line is variable. 
 For completeness we indicate in Fig.~\ref{het1rms} the positions
 of other highly ionized Fe species. There is marginal evidence for
 variability in [\ion{Fe}{vii}], but the S/N is too low to make
 any claims regarding the [\ion{Fe}{xiv}]$\lambda5303$ line. 
 It should be emphasized that the forbidden [\ion{Fe}{x}]$\lambda6375$ line
 is variable in Mrk\,110 while the permitted \ion{Fe}{ii} line blends 
 remained constant. The [\ion{Fe}{x}]$\lambda6375$ line and the 
 \ion{Fe}{ii} lines show the same intensity in the mean spectrum.

 The variability behaviour of the
 \ion{Fe}{ii} line blends in Seyfert galaxies is still poorly understood. 
 In some Seyfert\,1 galaxies the optical \ion{Fe}{ii} line blends
 are variable while in other galaxies no variations could be
 detected (Kollatschny et al. \cite{kollatschny00},
 Kollatschny \& Welsh \cite{kollatschny01}).
 There is evidence suggesting
 that the variability amplitude of 
 optical \ion{Fe}{ii} line blends in Seyfert 1 galaxies might be correlated
 with the emission line widths.
 This might be 
 an optical depth or obscurational effect,
 but this deserves considerably more detailed investigation.

 In Table~\ref{Heint} we list observed line ratios of the \ion{He}{i}
 singlet lines at $\lambda4922$, $\lambda5016$, and $\lambda6678$
 as well as of the triplet lines at $\lambda4471$ and $\lambda5876$.
%
%------------------------------------------------------------------------------
%
%\setcounter{table}{1}
 \begin{table}
 \caption{Observed and theoretical \ion{He}{i} line ratios: Internal line
 ratios of the singlet \ion{He}{i}$\lambda4922$, $\lambda5016$, 
 $\lambda6678$ lines, of the triplet \ion{He}{i}$\lambda5876$,
 $\lambda4471$ lines, and of the singlet \ion{He}{i}$\lambda5016$ to
 triplet \ion{He}{i}$\lambda4471$ lines.}
 \label{Heint}
 \begin{tabular}{lcc}
 \hline 
 \noalign{\smallskip}
  Line & Obs. flux ratio & Theor. flux ratio \\
 \noalign{\smallskip}
 (1) & (2) & (3) \\
 \noalign{\smallskip}
 \hline 
 \noalign{\smallskip}
\ion{He}{i}~$\lambda4922$/$\lambda5016$~ & 0.41 $\pm$ 0.2 & 0.45 \\ 
\ion{He}{i}~$\lambda6678$/$\lambda5016$~ & 1.36  $\pm$ 0.6 & 1.33 \\
 \noalign{\smallskip}
 \hline 
 \noalign{\smallskip}
\ion{He}{i}~$\lambda5876$/$\lambda4471$~  & 3.6  $\pm$ 0.3  & 2.97 \\
 \noalign{\smallskip}
 \hline 
 \noalign{\smallskip}
 \ion{He}{i}~$\lambda5016$/$\lambda4471$~  & 1.1  $\pm$ 0.4  & 0.58 \\
 \noalign{\smallskip}
 \hline \\
 \end{tabular}
 \end{table}
%
%----------------------------------------------------------------------------- 
%
 The line ratios have been determined from the rms spectrum
 (Fig.~\ref{het1rms}).
 The individual rms line intensities have errors of 5 to 40\%.
 The intrinsic line ratio \ion{He}{i}$\lambda5016$/$\lambda4471$~  might
 be lower by 10--20\% than the observed one due to the fact that the
 \ion{He}{i}$\lambda5016$ line
 is blended with the \ion{He}{i}$\lambda5048$ line.
 The rms spectrum is free of contamination of
 the constant narrow emission lines and \ion{Fe}{ii} line blends.
 Analysis of the difference spectrum (Fig.~\ref{het1rms})
 yields the same results within the errors.
 In the mean spectrum of Mkn\,110 (Fig.~\ref{het1mean})
 we could measure only the  \ion{He}{i}$\lambda5876$
 to \ion{He}{i}$\lambda4471$ line ratio. It agrees with that of the rms
 spectrum.
 Therefore it seems reliable that the mean to rms line ratios of the other  
 \ion{He}{i} lines show the same behavior.

 Theoretical \ion{He}{i} line ratios are listed in Table~\ref{Heint} column (3)
 for T~=~10,000 K and $n_{e}$~=~$10^{6}$ $cm^{-3}$
 (Benjamin et al.\ \cite{benjamin99}).
There are discrepancies of 30--40\% between theory and observations.
Much of this discrepancy might be attributed to the low 
density of the model calculations. The density in the broad-line region 
of Mrk\,110 might be 2-3 orders of magnitude higher.

The strong He emission detected in Mrk\,110 is similar to
the strong He emission seen in other accretion--powered sources,
such as the cataclysmic variables and X-ray binaries
(e.g. Warner \cite{warner95}).

 Further weak features in the rms spectrum could be attributed to
 \ion{He}{i}, \ion{He}{ii} and highly ionized [\ion{Fe}] lines. But a 
 detailed investigation of these lines is beyond the scope of this paper.

\subsection{The [\ion{O}{iii}]$\lambda$5007 wing}

The mean spectrum of Mrk\,110 (Fig.~\ref{het1mean}) shows broad extended
emission on the red side of the [\ion{O}{iii}]$\lambda$5007 line.
A thorough analysis of this spectral feature is important since
forbidden line emission of [\ion{O}{iii}]$\lambda$5007 
 from the BLR would allow a more
definite determination of gas density in this region.

In principle there are four possibilities to explain this broad emission: 
 \begin{enumerate}
  \item underlying \ion{Fe}{ii} line blends
  \item a strong extended red wing of the H$\beta$ line
  \item a broad red wing of the [\ion{O}{iii}]$\lambda$5007 line
  \item unresolved weak, narrow, forbidden lines
  or contribution from few weak, broad, permitted emission lines
 \end{enumerate}

The broad emission cannot be due to \ion{Fe}{ii} line blends because the 
feature remained present after subtraction of the scaled 
\ion{Fe}{ii} template spectrum (Fig.~\ref{het1fe}). Furthermore,
the \ion{Fe}{ii} lines did not vary
 during our observations and disappear in the rms-spectrum, but the
broad wing on the  [\ion{O}{iii}]$\lambda$5007 line is present
 in the rms spectrum (Fig.~\ref{het1rms}).

Sometimes a broad wing on the red side  
of the [\ion{O}{iii}]$\lambda$5007 line
has been assigned to a very broad red H$\beta$ wing as e.g. in
Akn\,120 (Kollatschny et al. \cite{kollatschny81}, Meyers \& Peterson
\cite{meyers85} and references therein).
This possibility can be ruled out 
because the line is relatively narrow. 
The broad component on the red side of 
[\ion{O}{iii}]$\lambda$5007 is clearly separated from H$\beta$.

The feature can not be a broad 
red wing of the [\ion{O}{iii}]$\lambda$5007 line
since there is no corresponding blue wing. Furthermore, the
 [\ion{O}{iii}]$\lambda$4959 and [\ion{O}{iii}]$\lambda$4363 lines 
do not have these wings (Fig.~\ref{het1rms}).

The feature cannot be due to constant NLR
emission because these lines would cancel out in the rms spectrum. The
most likely explanation is that the feature is due to broad
\ion{He}{i}$\lambda5016$ emission.

The \ion{He}{i}$\lambda6678$ line has been verified in some
 other high S/N spectra 
 of Seyfert galaxies (Filippenko \& Sargent \cite{filip85}).
 The \ion{He}{i}$\lambda5016$ line is more easily detected in the
 rms spectrum because it is blended with [\ion{O}{iii}]
 and the [\ion{O}{iii}] line cancels out
 in the rms spectrum.
 Indeed, there are indications in the
 published rms spectra of Seyfert
 galaxies indicating existing broad
 \ion{He}{i}$\lambda5016$ line emission
 as e.g. in 3C\,120, Mrk\,335, Mrk\,590 and Mrk\,817
 (Peterson \cite{peterson98}).  

In the next section we will demonstrate that
all \ion{He}{i} lines (including \ion{He}{i}$\lambda5016$)
are delayed by 10 -- 15 light days
with respect to continuum variations.
 This is a further demonstration that the red shoulder
 of the [\ion{O}{iii}] line is caused by \ion{He}{i} emission.

\subsection{CCF analysis and virial mass estimation of the central black hole}

The size and structure of a broad-line region in AGN can be estimated
from the cross-correlation function (CCF) of the light curve
of the ionizing continuum flux with the light curves
of the variable broad emission lines.

\begin{figure}
  \includegraphics[bb=40 60 570 780,width=63mm,angle=270]{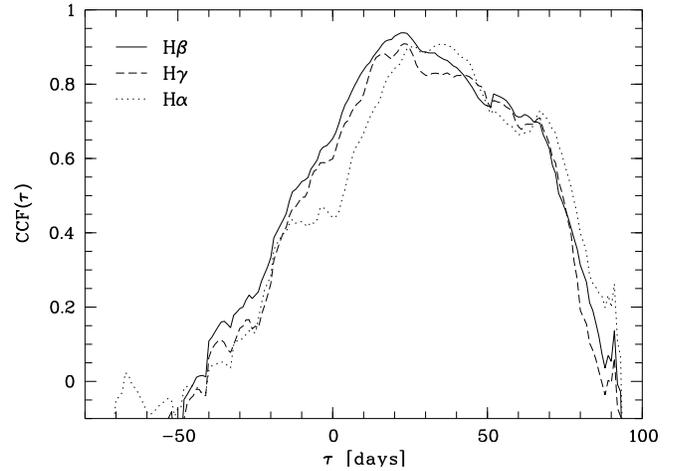}
  \caption{Cross-correlation functions CCF($\tau$) of the Balmer line light
           curves with the 5100\,\AA\ continuum light curve.}
  \label{ccfsh}
\end{figure}

\begin{figure}
  \includegraphics[bb=40 60 570 780,width=63mm,angle=270]{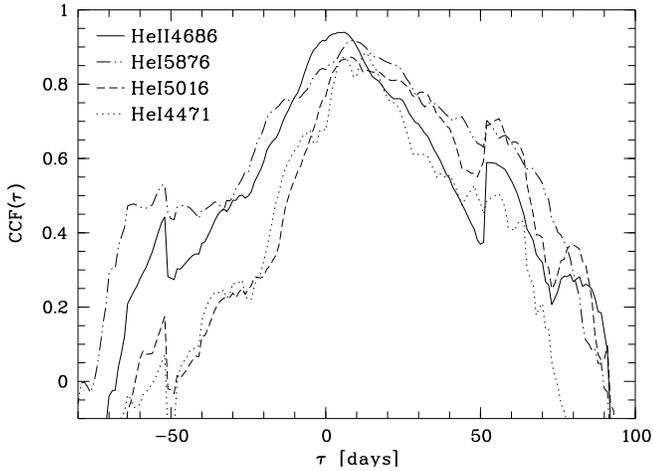}
  \caption{Cross-correlation functions CCF($\tau$) of the Helium line light
    curves with the 5100\,\AA\ continuum light curve.}
  \label{ccfshe}
\end{figure}

We cross-correlated the 5100\,\AA\ continuum light curve with all our emission
line light curves (Fig.\,3) using an interpolation cross-correlation function
 method (ICCF) described by Gaskell \& Peterson (\cite{gaskell87}).
The cross-correlation functions
of the Balmer lines (H$\gamma$, H$\beta$, and H$\alpha$) are plotted in 
Fig.~\ref{ccfsh}, those of HeII$\lambda$4686, HeI$\lambda$5876,
HeI$\lambda$5016 and HeI$\lambda$4471 in Fig.~\ref{ccfshe}.
%
%------------------------------------------------------------------------------
%
%
%------------------------------------------------------------------------------
%

The average interval between the observations was 7.3 days,
the median interval was 3.0 days.
The strong variability amplitudes in the continuum flux and in
the emission line intensities on time scales of weeks to months  
point to a very good sampling of the light curves.

We estimated the uncertainties in the cross-correlation results by
calculating the cross-correlation lags a large number of times
adding  random noise to our measured flux values as suggested by
Peterson et al. (\cite{peterson98b}).
 Furthermore,
the sampling uncertainties were estimated by considering different subsets
of our light curves
and repeating the cross-correlation calculations.
Typically we excluded
33\% of our spectra from the data set.
Finally we determined the error of the
lags by averaging and computing the standard deviation of the distribution
of centroid values. Centroids of the CCF, $\tau_{cent}$,
 were calculated using only the part of the CCF 
 above 85\% of the peak value.
% cross-correlation functions (e.g. Welsh \cite{welsh99}).
While determining accurate errors in CCF lags is difficult
(e.g. see Welsh \cite{welsh99}), the method we employed based on 
Peterson et al. (\cite{peterson98b}) should provide fairly reliable
error estimates.

In Table\,6 
we list our final cross-correlation results.
\begin{table}
\caption{Cross-Correlation lags of the emission lines
     with respect to the 5100\,\AA\ continuum.}
\begin{tabular}{lc}
\hline 
\noalign{\smallskip}
Line & \multicolumn{1}{c}{$\tau_{cent}$} \\
     & \multicolumn{1}{c}{[days]}\\
(1)  & \multicolumn{1}{c}{(2)}\\
\noalign{\smallskip}
\hline
\noalign{\smallskip}
HeII$\lambda 4686$~ &   $3.9^{+2.8}_{-0.7}$\\[.7ex]
HeI$\lambda 4471$   &   $11.1^{+6.0}_{-6.0}$\\[.7ex]
HeI$\lambda 5016$   &   $14.3^{+7.0}_{-7.0}$\\[.7ex]
HeI$\lambda 5876$~  &   $10.7^{+8.0}_{-6.0}$\\[.7ex]
H$\gamma$           &   $26.5^{+4.5}_{-4.7}$\\[.7ex]
H$\beta$            &   $24.2^{+3.7}_{-3.3}$\\[.7ex]
H$\alpha$           &   $32.3^{+4.3}_{-4.9}$\\

\noalign{\smallskip}
\hline 
\end{tabular}
\end{table}
Also, we calculated the cross-correlation lags of the
Balmer lines against each other,
and for the HeI lines as well.
 All three HeI lines show no internal lags
within the
uncertainties. The same holds for H$\beta$ and H$\gamma$. 
But the H$\alpha$ light curve is delayed with respect to the
H$\beta$ light curve by $7.5^{+2.}_{-2.}$ days. 
The same trend is seen in other Seyfert galaxies: the
H$\alpha$ rms line width is narrower than that of H$\beta$
(e.g. NGC\,4593 (Kollatschny \& Dietrich \cite{kollatschny97}))
and/or the H$\alpha$ response lags behind H$\beta$ (e.g. NGC\,3516
(Robinson \cite{robinson94})).
Calculations of the response functions for the broad emission
 lines in AGN show that the 
 H$\alpha$ line lags H$\beta$ 
 in most models (O'Brien et al. \cite{obrien94}).

The different delays of the HeII and HeI lines indicate an ionization
stratification structure of the BLR as seen in e.g. NGC\,5548 before
(e.g. Clavel et al. \cite{clavel91}).
 The fact that the HeI lines 
originate closer to the central source than the Balmer lines suggests
a density stratification in the BLR as well. Again the model calculations
of O'Brien et al. (\cite{obrien94}) confirm the observed structures
in the BLR.

All cross-correlation lags of our present variability campaign are smaller
by a factor of two than those determined for our 
long-term variability campaign from 1987 through 1995 (Bischoff \&
Kollatschny \cite{bischoff99}): The H$\beta$ lag is now
$24.2^{+3.7}_{-3.3}$ days instead of  $39.9^{+33.2}_{-9.5}$ days.
While this is formally only $<$\,1.7$\sigma$ discrepancy, we believe 
the difference to be real for the following reason:
It has been shown before as part of the NGC\,5548 variability campaign
that the characteristic radius of the BLR 
derived from the cross-correlation function 
depends on the duration and strength of the continuum outburst
(Dietrich \& Kollatschny \cite{dietrich95}).
During this HET variability campaign the luminosity of Mrk\,110
was in a all time low stage (see Fig.\,4).
Therefore, the radius of the BLR is likely to be smaller than
the long-term radius. 

It is possible to estimate the central mass in Mrk~110 from
the width of the broad emission line profiles (FWHM) under the assumption 
that the gas dynamics are dominated by the central massive object:
\[ M = \frac{3}{2} v^{2} G^{-1} R . \]
We presume that the characteristic velocity of the emission line
region is given by the FWHM of the rms profile and the characteristic 
distance R is given by the centroid of the corresponding cross-correlation
function (e.g. Koratkar \& Gaskell \cite{koratkar91}, Kollatschny \& Dietrich
\cite{kollatschny97}).

In Table\,7 we list our virial mass estimations of the central massive object
 in Mrk~110.
\begin{table}
\caption{Rms line widths (FWHM) of the strongest emission lines and
virial mass estimations of the central black hole.}
\begin{tabular}{lccc}
\hline 
\noalign{\smallskip}
Line & FWHM(rms) & $M$\\
     & [km s$^{-1}$] & [$10^7 M_{\odot}$]\\
(1)  & (2) & (3) \\ 
\noalign{\smallskip}
\hline
\noalign{\smallskip}
HeII$\lambda 4686$~ & 4444. $\pm$ 200. & $2.25^{+1.63}_{-0.45}$\\
HeI$\lambda 5876$~  & 2404. $\pm$ 100. & $1.81^{+1.36}_{-1.03}$\\
H$\beta$            & 1515. $\pm$ 100. & $1.63^{+0.33}_{-0.31}$\\
H$\alpha$           & 1315. $\pm$ 100. & $1.64^{+0.33}_{-0.35}$\\
\noalign{\smallskip}
\hline 
\noalign{\smallskip}
mean                &    & $1.83^{+0.54}_{-0.30}$\\
\noalign{\smallskip}
\hline
\end{tabular}
\end{table}
The best value considering all lines is
\[ M= 1.83^{+0.54}_{-0.30} \times 10^{7} M_{\odot} . \]
The virial mass estimations derived independently
from the different emission lines
agree within of 20\% with each other showing that uncertainties 
in our line measurements of this variability campaign
a relatively small. 

Peterson et al. (\cite{peterson98}) computed cross correlation lags
of 31.6 days (entire data set of 95 observations) and 19.5 days
(best subset of 14 observations only) for their Mrk\,110 variability
 campaign of H$\beta$. Their sampling was slightly worse than ours.
 They obtained a virial mass of 
$1.8 \times 10^{7} M_{\odot}$
 (H$\beta$ delay: 19.5 days, FWHM(H$\beta$):
2500 km s$^{-1}$). By correcting the FWHM of H$\beta$ to
1670 km s$^{-1}$ Wandel, Peterson, \& Malkan (\cite{wandel99}) 
computed a virial mass of $0.80 \times 10^{7} M_{\odot}$.
Their revised virial mass of $0.80 \times 10^{7} M_{\odot}$ is in good
agreement with our H$\beta$ virial mass of $1.63 \times 10^{7} M_{\odot}$,
especially considering that our formula for
computing the virial mass of the central object
yields masses 
 that are systematically higher
by a factor of two.
But one has to keep in mind that the derived central mass assumes
the very simple formula given above and systematic errors
as large as factors of several are possible.
Errors in echo mapping masses (e.g. Krolik \cite{krolik01}) due to
projection and/or geometry effects 
are not considered in this first order approximation.

\section{Conclusion}

We obtained very high S/N spectra with dense temporal sampling
during our variability campaign of the narrow-line Seyfert 1 galaxy Mrk\,110.
The homogeneous data set was obtained under identical instrumental conditions.
The central continuum flux was in a historically low stage  
during our observing campaign.

The main results of the present paper can be summarized as follows.
\begin{enumerate}
\item Considering the delays of the emission lines with respect to the
    continuum variations we verified  
    an ionization stratification in the BLR. The
    HeII$\lambda 4686$ line originates at a distance of 3.9 light days
    from the central ionizing source, the HeI lines at distances of 
    10--14 light days, and H$\beta$ and H$\alpha$
    at different distances of 24.2 and 32.3 light days respectively.
\item
 The forbidden [\ion{Fe}{x}]$\lambda6375$ line
 was variable in Mrk\,110 while the permitted \ion{Fe}{ii} line blends 
 remained constant.
\item
 We have shown that 
   the broad red wing of the [\ion{O}{iii}]$\lambda$5007 line is caused
   by the \ion{He}{i}$\lambda5016$ line only. 
   The observed relative intensities of the \ion{He}{i} lines
    in Mrk\,110 agree with theoretical model calculations.
\item  We derived virial masses of the central     
    black hole from the radial distances of different emission lines
    and their line widths. 
    The calculated central masses agree within 20\%.
     We determined a central mass of
\[ M= 1.83^{+0.54}_{-0.30} \times 10^{7} M_{\odot} . \]
\end{enumerate}

\begin{acknowledgements}
      WK thanks the UT Astronomy Department for warm hospitality during
      his visit. We thank the Resident Astronomers Matthew Shetrone and 
      Grant Hill and the HET staff.
      The Marcario Low Resolution Spectrograph is a joint project of the
      Hobby-Eberly Telescope partnership and the Instituto de Astronomia
      de la Universidad Nacional Autonoma de Mexico.
      Part of this work has been supported by the
      \emph{Deut\-sche For\-schungs\-ge\-mein\-schaft, DFG\/} grant
      KO 857/24 and DARA, and also the National Science Foundation under Grants
      AST-0086692 and INT-0049045.
\end{acknowledgements}

\newpage
%\begin{landscape}
\setcounter{table}{2}
\begin{table}
\caption{Continuum and integrated line fluxes}
\begin{tabular}{crrrrrrrrrr}
\noalign{\smallskip}
\hline 
\noalign{\smallskip}
Julian Date & 5100\,\AA & H$\alpha$ & H$\beta$ & H$\gamma$ & HeII$\lambda 4686$
 & HeI$\lambda 5876$ & HeI$\lambda 5016$ & HeI$\lambda 4471$ \\
2\,400\,000+\\
(1) & (2) & (3) & (4) & (5) & (6) & (7) & (8) & (9) \\ 
\noalign{\smallskip}
\hline
\noalign{\smallskip}
51495.94&  1.544 $\pm\ $0.025 &  1158.8 $\pm\ $25. &  226.4 $\pm\ $ 4. &  65.67
 $\pm\ $ 4.&  40.24 $\pm\ $2.5 &  49.38 $\pm\ $3.0 &  19.34 $\pm\ $2.0 &  10.38
 $\pm\ $1.5 & \\
51497.91&  1.559 $\pm\ $0.025 &  1184.6 $\pm\ $25. &  229.4 $\pm\ $ 4. &  67.97
 $\pm\ $ 4.&  39.42 $\pm\ $2.5 &  46.87 $\pm\ $3.0 &  18.69 $\pm\ $2.0 &   8.83
 $\pm\ $1.5 & \\   
51500.91&  1.646 $\pm\ $0.025 &  1173.6 $\pm\ $25. &  231.8 $\pm\ $ 4. &  76.86
 $\pm\ $ 4.&  41.09 $\pm\ $2.5 &  49.50 $\pm\ $3.0 &  18.41 $\pm\ $2.0 &  10.60
 $\pm\ $1.5 & \\   
51518.89&  1.920 $\pm\ $0.010 &  1112.2 $\pm\ $ 5. &  237.0 $\pm\ $ 2. &  78.89
 $\pm\ $ 2.&  55.72 $\pm\ $1.5 &  47.59 $\pm\ $1.5 &  21.31 $\pm\ $1.0 &  12.84
 $\pm\ $1.0 & \\   
51520.87&  1.917 $\pm\ $0.011 &  1082.1 $\pm\ $ 5. &  232.7 $\pm\ $ 2. &  72.55
 $\pm\ $ 2.&  51.98 $\pm\ $1.5 &  46.66 $\pm\ $1.5 &  21.56 $\pm\ $1.0 &  12.38
 $\pm\ $1.0 & \\   
51522.88&  1.937 $\pm\ $0.018 &  1040.2 $\pm\ $ 5. &  235.6 $\pm\ $ 2. &  74.01
 $\pm\ $ 2.&  53.54 $\pm\ $1.5 &  46.55 $\pm\ $1.5 &  22.59 $\pm\ $1.0 &  12.34
 $\pm\ $1.0 & \\   
51525.84&  1.821 $\pm\ $0.012 &  1197.6 $\pm\ $ 5. &  258.0 $\pm\ $ 2. &  82.39
 $\pm\ $ 2.&  62.04 $\pm\ $1.5 &  52.20 $\pm\ $1.5 &  24.78 $\pm\ $1.0 &  14.38
 $\pm\ $1.0 & \\   
51528.84&  1.858 $\pm\ $0.008 &  1205.2 $\pm\ $ 5. &  252.7 $\pm\ $ 2. &  82.30
 $\pm\ $ 2.&  59.13 $\pm\ $1.5 &  53.08 $\pm\ $1.5 &  22.92 $\pm\ $1.0 &  13.44
 $\pm\ $1.0 & \\   
51547.80&  2.155 $\pm\ $0.011 &  1204.2 $\pm\ $ 5. &  265.2 $\pm\ $ 2. &  84.79
 $\pm\ $ 2.&  83.81 $\pm\ $1.5 &  54.39 $\pm\ $1.5 &  26.40 $\pm\ $1.0 &  14.06
 $\pm\ $1.0 & \\   
51584.72&  1.413 $\pm\ $0.013 &  1252.4 $\pm\ $ 5. &  270.0 $\pm\ $ 2. &  92.85
 $\pm\ $ 2.&  39.50 $\pm\ $1.5 &  45.21 $\pm\ $1.5 &  21.91 $\pm\ $1.0 &  13.18
 $\pm\ $1.0 & \\   
51586.71&  1.392 $\pm\ $0.007 &  1261.2 $\pm\ $ 5. &  262.9 $\pm\ $ 2. &  84.75
 $\pm\ $ 2.&  35.42 $\pm\ $1.5 &  45.65 $\pm\ $1.5 &  21.16 $\pm\ $1.0 &  11.97
 $\pm\ $1.0 & \\   
51595.88&  1.546 $\pm\ $0.050 &  1194.5 $\pm\ $10. &  235.1 $\pm\ $ 4. &  86.77
 $\pm\ $ 4.&  37.16 $\pm\ $3.0 &  43.16 $\pm\ $3.0 &  19.64 $\pm\ $2.0 &  12.32
 $\pm\ $3.0 & \\   
51598.86&  1.634 $\pm\ $0.025 &  1125.3 $\pm\ $ 5. &  239.2 $\pm\ $ 2. &  73.89
 $\pm\ $ 2.&  47.25 $\pm\ $1.5 &  44.83 $\pm\ $1.5 &  23.02 $\pm\ $1.0 &  10.40
 $\pm\ $1.0 & \\   
51605.83&  1.405 $\pm\ $0.014 &  1143.1 $\pm\ $ 5. &  238.2 $\pm\ $ 2. &  82.08
 $\pm\ $ 2.&  44.21 $\pm\ $1.5 &  46.37 $\pm\ $1.5 &  21.15 $\pm\ $1.0 &  12.69
 $\pm\ $1.0 & \\   
51608.62&  1.350 $\pm\ $0.015 &  1128.5 $\pm\ $ 5. &  230.1 $\pm\ $ 2. &  68.28
 $\pm\ $ 2.&  37.87 $\pm\ $1.5 &  44.14 $\pm\ $1.5 &  21.17 $\pm\ $1.0 &  10.97
 $\pm\ $1.0 & \\   
51611.62&  1.364 $\pm\ $0.017 &  1091.8 $\pm\ $ 5. &  229.8 $\pm\ $ 2. &  73.56
 $\pm\ $ 2.&  30.40 $\pm\ $1.5 &  39.32 $\pm\ $1.5 &  20.89 $\pm\ $1.0 &  11.38
 $\pm\ $1.0 & \\   
51614.63&  1.088 $\pm\ $0.011 &  1140.5 $\pm\ $ 5. &  234.6 $\pm\ $ 2. &  79.75
 $\pm\ $ 2.&  29.53 $\pm\ $1.5 &  37.42 $\pm\ $1.5 &  18.56 $\pm\ $1.0 &  10.40
 $\pm\ $1.0 & \\   
51629.76&  1.076 $\pm\ $0.013 &  1076.2 $\pm\ $ 5. &  201.3 $\pm\ $ 2. &  52.80
 $\pm\ $ 2.&  28.74 $\pm\ $1.5 &  33.60 $\pm\ $1.5 &  18.16 $\pm\ $1.0 &   8.46
 $\pm\ $1.0 & \\   
51637.77&  1.042 $\pm\ $0.009 &  1066.7 $\pm\ $ 5. &  191.6 $\pm\ $ 2. &  52.53
 $\pm\ $ 2.&  27.36 $\pm\ $1.5 &  31.59 $\pm\ $1.5 &  17.31 $\pm\ $1.0 &   8.98
 $\pm\ $1.0 & \\   
51645.73&  1.156 $\pm\ $0.015 &  1009.3 $\pm\ $ 5. &  182.3 $\pm\ $ 2. &  52.33
 $\pm\ $ 2.&  21.72 $\pm\ $1.5 &  33.91 $\pm\ $1.5 &  15.91 $\pm\ $1.0 &   9.30
 $\pm\ $1.0 & \\   
51658.70&  1.385 $\pm\ $0.019 &   839.7 $\pm\ $ 5. &  180.3 $\pm\ $ 2. &  47.16
 $\pm\ $ 2.&  29.74 $\pm\ $1.5 &  29.45 $\pm\ $1.5 &  19.68 $\pm\ $1.0 &  10.76
 $\pm\ $1.0 & \\   
51663.68&  1.263 $\pm\ $0.042 &   950.7 $\pm\ $10. &  184.4 $\pm\ $ 4. &  50.06
 $\pm\ $ 3.&  28.05 $\pm\ $2.5 &  33.09 $\pm\ $2.5 &  17.37 $\pm\ $2.0 &   9.93
 $\pm\ $2.0 & \\   
51664.66&  1.209 $\pm\ $0.009 &   973.2 $\pm\ $ 5. &  185.7 $\pm\ $ 2. &  57.60
 $\pm\ $ 3.&  30.82 $\pm\ $1.5 &  33.16 $\pm\ $1.5 &  16.59 $\pm\ $1.0 &   9.41
 $\pm\ $1.0 & \\   
51670.70&  1.329 $\pm\ $0.021 &   985.0 $\pm\ $ 8. &  179.4 $\pm\ $ 3. &  54.45
 $\pm\ $ 3.&  23.75 $\pm\ $2.5 &  36.99 $\pm\ $2.5 &  14.85 $\pm\ $2.0 &  10.59
 $\pm\ $2.0 & \\   
51673.69&  1.110 $\pm\ $0.017 &   917.5 $\pm\ $ 5. &  190.0 $\pm\ $ 2. &  54.93
 $\pm\ $ 2.&  22.82 $\pm\ $1.5 &  31.06 $\pm\ $1.5 &  18.81 $\pm\ $1.0 &  12.73
 $\pm\ $1.0 & \\
51678.64&  1.108 $\pm\ $0.013 &   982.6 $\pm\ $ 5. &  188.3 $\pm\ $ 2. &  53.61
 $\pm\ $ 2.&  25.95 $\pm\ $1.5 &  34.69 $\pm\ $1.5 &  16.72 $\pm\ $1.0 &   9.14
 $\pm\ $1.0 & \\
\noalign{\smallskip}
\hline 
\noalign{\smallskip}
\end{tabular}

Continuum flux (2) in 10$^{-15}$\,erg\,sec$^{-1}$\,cm$^{-2}$\,\AA$^{-1}$.\\
Line fluxes (3) - (9) in 10$^{-15}$\,erg\,sec$^{-1}$\,cm$^{-2}$.
\end{table}
%\end{landscape}

\end{document}